# Physical model for vaporization

Jozsef Garai

*Department of Mechanical and Materials Engineering, Florida International University, University Park, VH 183, Miami, FL 33199*

**Abstract**

Based on two assumptions, the surface layer is flexible, and the internal energy of the latent heat of vaporization is completely utilized by the atoms for overcoming on the surface resistance of the liquid, the enthalpy of vaporization was calculated for 45 elements. The theoretical values were tested against experiments with positive result.

## 1. Introduction

The enthalpy of vaporization is an extremely important physical process with many applications to physics, chemistry, and biology. Thermodynamic defines the enthalpy of vaporization $(\Delta_v H)$ as the energy that has to be supplied to the system in order to complete the liquid-vapor phase transformation. The energy is absorbed at constant pressure and temperature. The absorbed energy not only increases the internal energy of the system (U) but also used for the external work of the expansion (w). The enthalpy of vaporization is then

$$\Delta_v H = \Delta_v U + \Delta_v w \quad (1)$$

The work of the expansion at vaporization is

$$\Delta_v w = P\left(V_V - V_L\right) \quad (2)$$

where p is the pressure, $V_V$ is the volume of the vapor, and $V_L$ is the volume of the liquid.

Several empirical and semi-empirical relationships are known for calculating the enthalpy of vaporization [1-16]. Even though there is no consensus on the exact physics, there is a general agreement that the surface energy must be an important part of the enthalpy of vaporization. The vaporization diminishes the surface energy of the liquid; thus this energy must be supplied to the system. Additionally it is well established that both enthalpy of vaporization and surface tension [$\gamma_{LV}$] are function of temperature.

Laplace in his theory suggested that the ratio of the molar total surface energy and the molar internal latent heat is constant with respect to the temperature. Weisskopf [12, 13] determined the total surface area by slicing the substance at the linear dimension of one molecule and summing these surfaces. The enthalpy of vaporization was calculated by multiplying the total surface area with the surface tension. The model was evaluated by comparing the theoretical atomic diameters to the actual ones

$$d = \frac{6\gamma_{LV}}{\varepsilon_B} \quad (3)$$

where $\gamma_{LV}$ is the surface tension at the boiling temperature, and $\varepsilon_B$ is the binding energy of the substance contained in one cubic centimeter volume. The theoretical values of the investigated ten substances agreed reasonable well with experiments.

Agrawal and Menon [10] proposed that the atomic layers are removed one by one from the liquid. They calculated the area of one gram liquid as:

$$A = \frac{1}{d\rho_L} \quad (4)$$

where d is the average distance between the molecules and $\rho_L$ is the density of the liquid. The internal energy required to remove all of the layers is then

$$\Delta U_\gamma = \gamma_{LV} A = \frac{\gamma_{LV}}{d\rho_L}. \quad (5)$$

The enthalpy of vaporization was calculated by summing the internal energy [Eq.(5)] and the work required for the volume expansion at the boiling temperature [Eq.(2)]. The calculated enthalpy of vaporization values were compared to experiments of Li, Na, K, Rb, Cs, and water. In order to reproduce the experiments the introduction of a multiplier [f] was necessary. The value of the multiplier is between 4.3 and 7.1.

Linear relationship between the enthalpy of vaporization and the product of the surface tension and the square of the radius of the molecule were reported by Keeney and Heicklen [15] as

$$\Delta_V H = k\gamma_{LV_c}[T] r_c^2 [T] + k' \quad (6)$$

where $r_c$ is the radius of molecule C, k and k' are constant, independent of temperature.

In this study a physical explanation for vaporization is proposed. The model allows calculating the multiplier [f] for Eqs. (5) and the constants for Eq. (6) from theory. The calculated enthalpy of vaporization of monoatomic liquids is tested against experiments.

## 2. Proposed model

The surface layer of a liquid is flexible. If an inside atom hits the surface layer then the first reaction is that the surface absorbs the energy through deformation. The maximum resistance of the deformed surface is achieved when the center of the outgoing atom reaches the plane of the surface as shown in Fig. 1. Beyond that maximum resistance the area of the surface starts to decrease and eventually leads to the detachment of the atom from the liquid. The extra energy required for an atom to escape from the liquid, is equal with the maximum surface area resistance. The maximum surface area [A] around an atom is approximated as:

$$A = 2\pi(2r)^2 = 8\pi r^2 \quad (7)$$

by assuming that the surface of the liquid contains one atomic layer and that the liquid is monoatomic [Fig. 1.]. In Eq. (7) r is the hard sphere radius of the atoms. The maximum surface area for molecular liquids is more complicated and not investigated here. Multiplying the maximum surface area of Eq. (7) with the number of moles [n], Avogadro's number [ $N_A$ ], and the surface tension at the boiling temperature gives the internal energy required for the vaporization of the bulk liquid atoms.

$$\Delta_V U = 8n\pi N_A \gamma_{LV} r^2 \tag{8}$$

Equation (8) can be written in a more general form as:

$$\Delta_V U = c\gamma_{LV} \tag{9}$$

where c is a constant, characteristic of the substance. The enthalpy of vaporization is the sum of the internal energy Eq. (8) and the external work Eq. (2). Assuming that the ideal gas law is valid and the volume of the liquid is negligible then the extension work can be calculated as:

$$\Delta_v w = PV_{V(T_b)} = nRT_b \tag{10}$$

where R is the universal gas constant, and $T_b$ is the boiling temperature in Kelvin. The enthalpy of vaporization is then

$$\Delta_v H = n\left[2\pi N_A \gamma_{LV} (2r)^2 + RT_b\right] . \tag{11}$$

Equation (9) predicts a linear relationship between the internal energy and the surface tension regardless of the temperature which is consistent with previous investigations [15, 17-19]. Using the experiments of water this linear relationship is shown on Fig. 2. Experiments up to 28 bar pressure are used because equations (10) can not be used for high pressures since the ideal gas law is not valid for highly dense gasses.

## 3. Testing the model against experiments

The theoretical relationship in Eq. (11) was tested against the experimental data of monoatomic liquids. The physical properties of the 45 elements used for the investigation are listed in Table 1. Using the experimental enthalpy of vaporization values, the radius of the surface area was determined in atomic radius units. The calculated values are reasonably close to the theoretical value of 2r with the exceptions of the elements Si, Ge, and Sn in group 14. The surface radius for these elements is between 2.62-2.88r. Possible explanation for this irregularity could be that the surface of these liquids contains two atomic layers and the maximum surface area could be approximated as:

$$A = 2\pi(3r)^2 = 18\pi r^2 . \tag{12}$$

The residuals, the difference between the observed and calculated values are plotted against the atomic radius and the surface tension and given in Fig. 2. Investigating the differences between the calculated and the experimental enthalpy of vaporizations two characteristics has been identified. One, the average of the calculated enthalpy of vaporization is about 10% higher than the experiments, and two elements of the different groups of the periodic table shows small but systematic deviations. Explanations for these characteristics are proposed.

The surface tensions used for the calculations, in many cases, were not measured at boiling but lower temperatures. These higher surface tension values overestimate the enthalpy of vaporization. The other contribution comes from the hardball representation of the atoms which can also explain the observed systematic differences between the different groups of the elements. The electron shell of the atoms is compressed when they break through the surface. This compression results in smaller surface area than calculated by Eq. (7) and smaller enthalpy than Eq. (11). The hardball approximation used in the model; therefore, overestimates the enthalpy. The deviation is the highest for elements with weak electron shell.

It should also be noted that not all atomic compounds evaporate to produce vapor system containing monoatomic. Alkali metals are known for diverting interatomic forces upon vaporization. The overlapping of the atoms in a polyatomic formation also reduces the surface area [Eq. (7)] and the enthalpy of vaporization [Eq. (11)].

Quantum effects due to mass, size, and interaction of energy can also result in deviation from the classical approach [34-35] used in this study. For heavy elements the deviation from classical result is negligible, however, for light elements, $D_2$, $H_2$, He, and for $N_2$, Ne, Ar quantum effects must be considered [34]. Deviation from classical description has also been reported for Li and in some extent for Na [36]. The quantum effect alters the surface tension [37]; therefore, the presented classical approach is not applicable to liquids with sizable quantum effect.

## 4. Conclusions

Physical model for the latent heat of vaporization is proposed. The model assumes that the surface layer is flexible and that the internal energy of the latent heat of vaporization is completely utilized by the atoms for overcoming on the surface resistance of the liquid. Based on these assumptions the energy required for the vaporization of the liquid can be calculated. The theoretical values of the enthalpy of vaporization were calculated for 45 elements from first principles. The calculated values generally agree well with experiments. The small deviations between the calculated and experimental values result from the lower temperature surface tensions used in the calculations, the hardball sphere approximation of the atoms, the non-monoatomic evaporation and quantum effects.

It is concluded that, with the exception of quantum liquids, the proposed classical model correctly describes the physical process of vaporization.

**Acknowledgement**

I would like to thank to Jeffrey Joens for reading the manuscript and making comments and the thoughtful suggestions of the anonymous reviewer. This research was supported by Florida International University Dissertation Year Fellowship and NSF EAR 0711321.

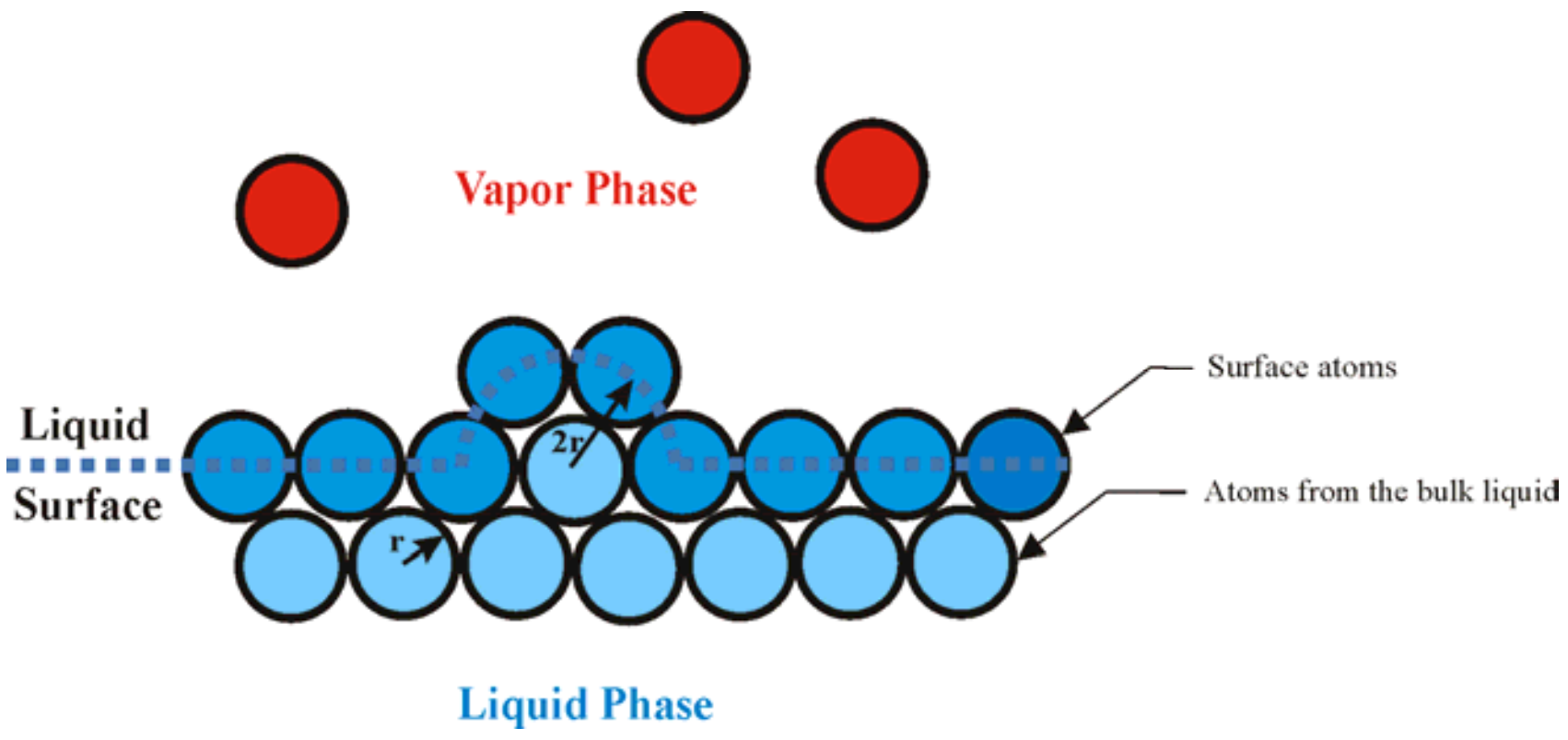

**Figure 1.** Schematic cross section of the proposed vaporization model for monoatomic liquids with one surface layer.

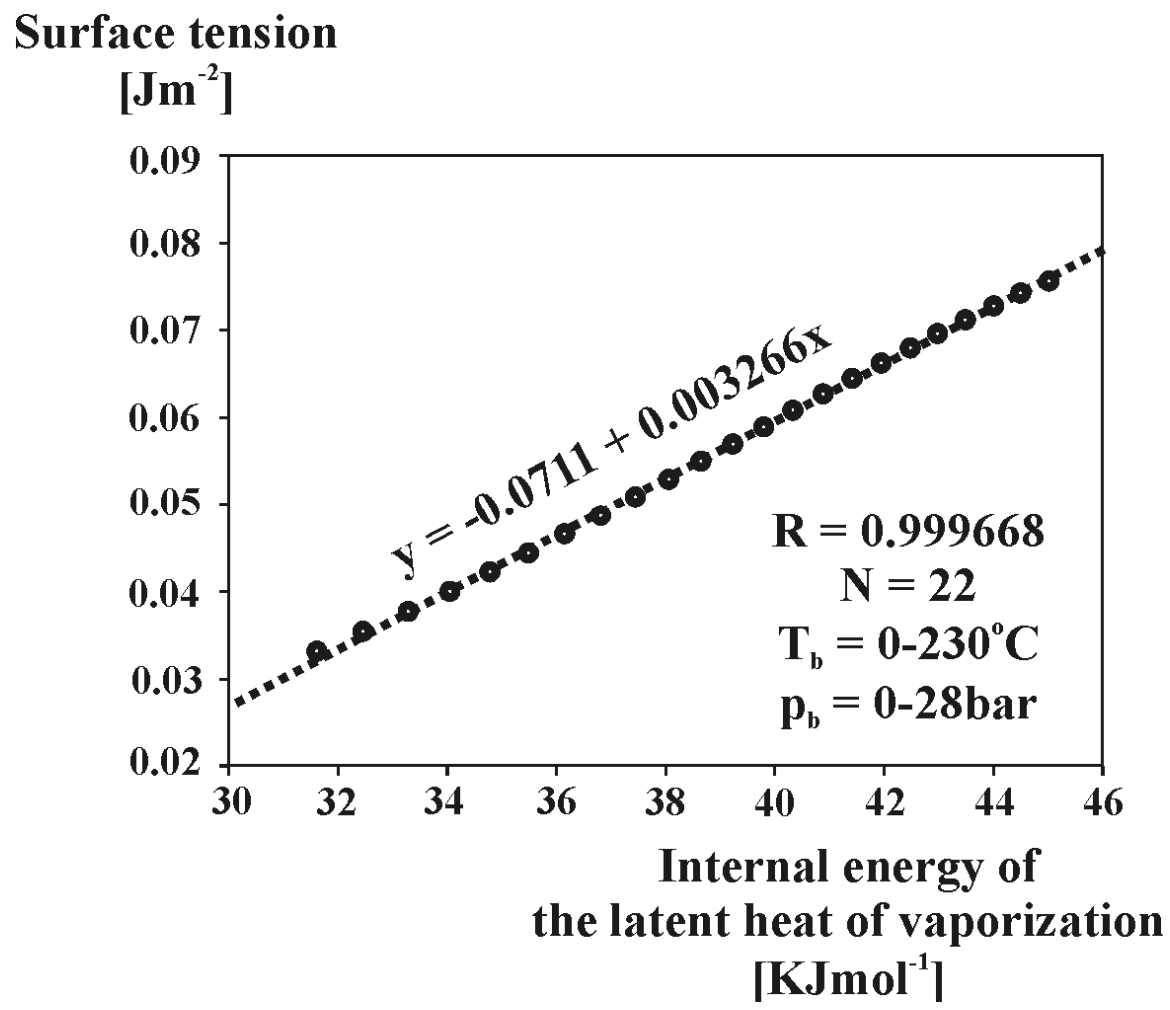

**Figure 2.** The surface tension is plotted against the internal energy of the latent heat of vaporization from experiments of water [17-19]. The experiments in the 0-28 bar pressure region are consistent with the theoretically predicted linear correlation. R is the correlation coefficient and N is the number of experiments.

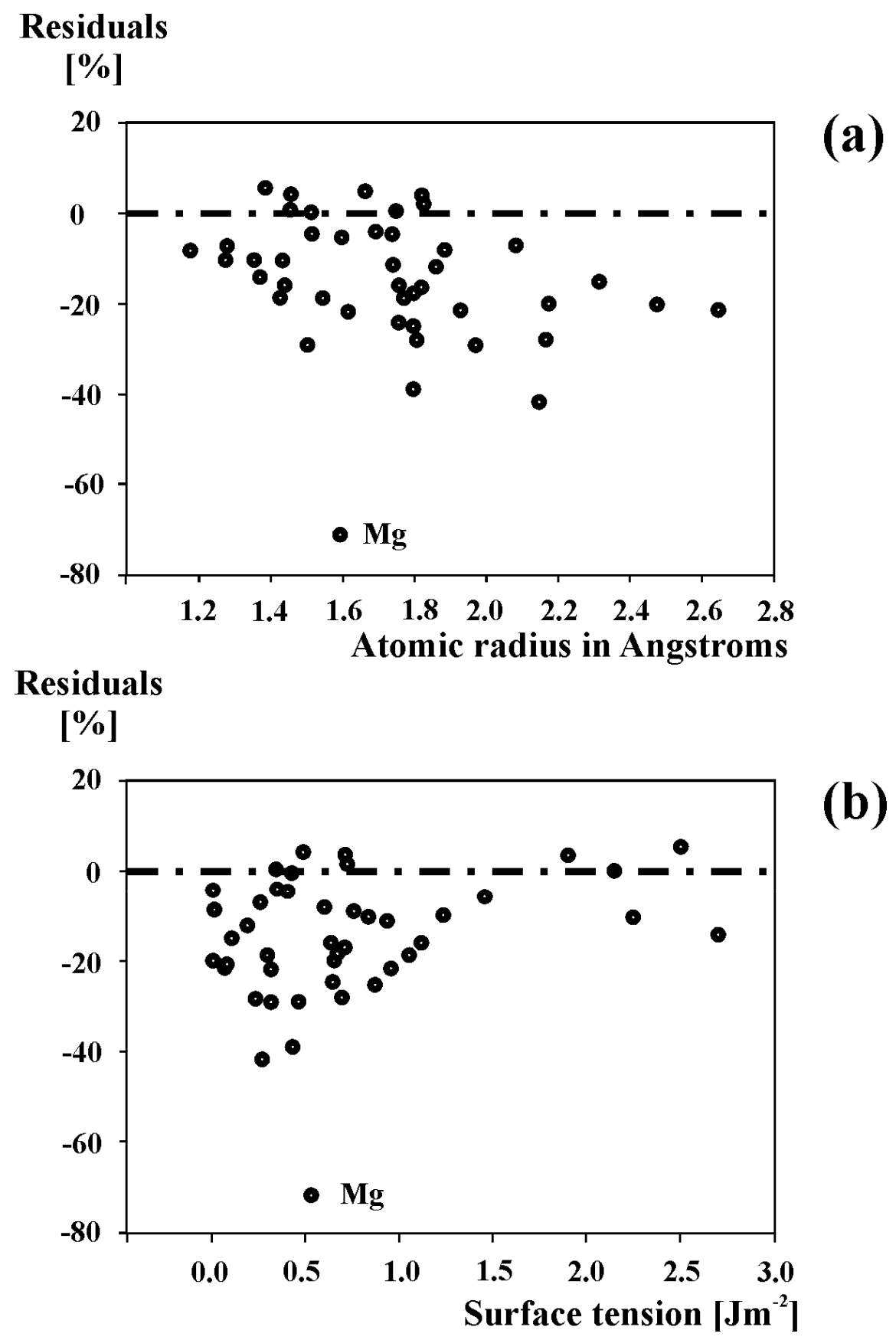

**Figure 3.** The residuals of the enthalpy of vaporization plotted against. (a) atomic radius. (b) surface tension.

**Table 1.** The physical parameters of the elements used for the investigation and the calculated enthalpies for the vaporization.

| Group | Atomic Number | Element | Atomic Radius [20] [Å] | Surface Tension $[J/m^2]$ | Boiling Temp. [20] [K] | Exp. Work [KJ] | Calc. rad. for max. surface resistance in atomic r | Enthalpy of Evaporation [KJ] Exp. [20] | Enthalpy of Evaporation [KJ] Calc. | Ratio | Average Ratio |
|---|---|---|---|---|---|---|---|---|---|---|---|
| | 3 | Li | 1.52 | 0.398 [22] | 1620 | 13.47 | 1.95 | 145.9 | 152.6 | 0.96 | |
| | 11 | Na | 1.86 | 0.189 [23] | 1156 | 9.61 | 1.88 | 97.0 | 108.6 | 0.89 | |
| 1 | 19 | K | 2.32 | 0.102 [23] | 1047 | 8.71 | 1.85 | 79.9 | 91.9 | 0.87 | 0.87 |
| | 37 | Rb | 2.48 | 0.085 [22] | 961 | 7.99 | 1.80 | 72.2 | 87.1 | 0.83 | |
| | 55 | Cs | 2.65 | 0.070 [22] | 951 | 7.91 | 1.79 | 67.7 | 82.3 | 0.82 | |
| | 12 | Mg | 1.60 | 0.516*[24, 25] | 1380 | 11.48 | 1.50 | 127.4 | 211.3 | 0.60 | |
| 2 | 20 | Ca | 1.97 | 0.297*[24, 25] | 1757 | 14.61 | 1.74 | 153.6 | 189.2 | 0.81 | 0.76 |
| | 38 | Sr | 2.15 | 0.250*[24, 25] | 1656 | 13.77 | 1.65 | 144.0 | 188.8 | 0.76 | |
| | 56 | Ba | 2.17 | 0.209*[24, 25] | 1913 | 15.91 | 1.74 | 142.0 | 164.9 | 0.86 | |
| | 21 | Sc | 1.62 | 0.954[20] | 3109 | 25.85 | 1.80 | 332.7 | 404.8 | 0.82 | |
| 3 | 39 | Y | 1.80 | 0.871[20] | 3618 | 30.08 | 1.77 | 365.0 | 457.2 | 0.80 | 0.88 |
| | 57 | La | 1.83 | 0.718[20] | 3737 | 31.07 | 2.02 | 402.1 | 395.0 | 1.021 | |
| 4 | 40 | Zr | 1.60 | 1.463 [26] | 4650 | 38.66 | 1.94 | 573.0 | 605.5 | 0.95 | 0.95 |
| 5 | 41 | Nb | 1.46 | 1.900 [27] | 5200 | 43.24 | 2.04 | 682.0 | 656.2 | 1.04 | 1.02 |
| | 73 | Ta | 1.46 | 2.150 [27] | 5698 | 47.38 | 2.00 | 743.0 | 741.0 | 1.00 | |
| 6 | 74 | W | 1.39 | 2.500 [27] | 5933 | 49.33 | 2.06 | 824.0 | 780.4 | 1.06 | 1.06 |
| 7 | 75 | Re | 1.37 | 2.700 [27] | 5900 | 49.06 | 1.86 | 715.0 | 816.1 | 0.88 | 0.88 |
| 9 | 77 | Ir | 1.36 | 2.250 [27] | 4800 | 39.91 | 1.90 | 604.0 | 665.2 | 0.91 | 0.91 |
| | 29 | Cu | 1.28 | 0.986*[28] | 2840 | 23.61 | 1.90 | 300.4 | 268.1 | 1.12 | |
| 11 | 47 | Ag | 1.44 | 0.640*[28] | 2485 | 20.66 | 1.90 | 258.0 | 221.7 | 1.16 | 1.14 |
| | 79 | Au | 1.44 | 0.899*[29] | 3129 | 26.02 | 1.84 | 324.0 | 308.2 | 1.02 | |
| 12 | 80 | Hg | 1.51 | 0.424 [30] | 630 | 5.24 | 2.00 | 151.0 | 151.6 | 1.00 | 1.00 |
| | 13 | Al | 1.43 | 1.050 [22] | 2740 | 22.78 | 1.82 | 293.0 | 348.2 | 0.84 | |
| 13 | 49 | In | 1.67 | 0.391*[28] | 2273 | 18.90 | 2.05 | 231.8 | 183.9 | 1.26 | 1.04 |
| | 81 | Tl | 1.70 | 0.331*[24, 25] | 1730 | 14.39 | 1.85 | 164.0 | 159.0 | 1.04 | |
| | 14 | Si | 1.18 | 0.760 [31, 32] | 3538 | 29.42 | 2.87 | 359.0 | 389.8 | 0.92 [dsl] | |
| 14 | 32 | Ge | 1.28 | 0.600[31, 32] | 3106 | 25.83 | 2.88 | 334.0 | 360.6 | 0.93 [dsl] | 0.95 |
| | 50 | Sn | 1.51 | 0.415*[28] | 2543 | 21.15 | 2.62 | 295.8 | 343.2 | 0.86 [dsl] | |
| | 82 | Pb | 1.75 | 0.310*[24, 25] | 2013 | 16.74 | 2.00 | 177.7 | 160.6 | 1.11 | |
| 15 | 83 | Bi | 1.55 | 0.268*[24, 25] | 1833 | 15.24 | 1.81 | 104.8 | 112.4 | 0.93 | 0.93 |
| | 18 | Ar | 1.74 [21] | 0.013 [33] | 87 | 0.73 | 1.95 | 6.4 | 6.7 | 0.96 | |
| 18 | 36 | Kr | 1.89 [21] | 0.016 [33] | 120 | 1.00 | 1.91 | 9.1 | 9.9 | 0.92 | 0.90 |
| | 54 | Xe | 2.18 [21] | 0.019 [33] | 167 | 1.38 | 1.81 | 12.6 | 15.2 | 0.83 | |
| | 58 | Ce | 1.82 | 0.706 [20] | 3716 | 30.90 | 2.04 | 398.0 | 384.1 | 1.04 | |
| | 59 | Pr | 1.82 | 0.707 [20] | 3793 | 31.54 | 1.83 | 331.0 | 387.5 | 0.85 | |
| | 60 | Nd | 1.81 | 0.687 [20] | 3347 | 27.83 | 1.75 | 289.0 | 370.0 | 0.78 | |
| | 62 | Sm | 1.80 | 0.431 [20] | 2067 | 17.19 | 1.67 | 165.0 | 229.5 | 0.72 | |
| | 63 | Eu | 2.08 | 0.264 [20] | 1802 | 14.98 | 1.93 | 176.0 | 188.5 | 0.93 | |
| L | 64 | Gd | 1.80 | 0.664 [20] | 3546 | 29.48 | 1.82 | 301.3 | 356.5 | 0.85 | 0.85 |
| | 65 | Tb | 1.77 | 0.669 [20] | 3503 | 29.13 | 1.82 | 293.0 | 347.4 | 0.84 | |
| | 66 | Dy | 1.78 | 0.648 [20] | 2840 | 23.61 | 1.82 | 280.0 | 334.7 | 0.84 | |
| | 67 | Ho | 1.76 | 0.650 [20] | 2973 | 24.72 | 1.77 | 265.0 | 330.2 | 0.80 | |
| | 68 | Er | 1.76 | 0.637 [20] | 3141 | 26.12 | 1.84 | 280.0 | 325.1 | 0.86 | |
| | 70 | Yb | 1.93 | 0.320 [20] | 1469 | 12.22 | 1.80 | 159.0 | 193.2 | 0.82 | |
| | 71 | Lu | 1.74 | 0.940 [20] | 3675 | 30.56 | 1.89 | 414.0 | 460.3 | 0.90 | |

*surface tension at boiling temperature calculated from the temperature coefficients of the surface tension (The rest of the listed surface tensions are given at temperatures close to the melting temperature.)
[dsl] double surface layer (explanation is in the text)